\documentclass{appolb}
\usepackage{graphicx}
\newcommand \Pomeron {I\!\!P}
\begin{document}
\title{SUPPRESSION OF DIFFRACTION IN DIS ON NUCLEI AND DYNAMICAL MECHANISM OF LEADING TWIST
NUCLEAR SHADOWING
\thanks{Presented at Diffraction and Low-$x$ 2024, Palermo, Italy, September 8-14, 2024}%
}
\author{V. Guzey
\address{University of Jyvaskyla, Department of Physics, P.O. Box 35, FI-40014 University
of Jyvaskyla, Finland and Helsinki Institute of Physics, P.O. Box 64, FI-00014
University of Helsinki, Finland}
}
\maketitle
\begin{abstract}

Using the leading twist approach (LTA) to nuclear shadowing (NS), we calculate 
the ratio of the diffractive-to-total DIS cross sections $R_{\rm diff/tot}$ for a heavy nucleus and proton
and confirm that $R_{\rm diff/tot} \approx 0.5-1$ in contrast with $R_{\rm diff/tot} \approx 1.5-2$ in the dipole model.
We show that the magnitude of $R_{\rm diff/tot}$ is controlled by an effective dipole cross section so that 
$R_{\rm diff/tot} \to 1$ in the black disk limit. We also argue that strong leading twist NS as well as the dilute nuclear density
deplete nuclear enhancement of the saturation scale leading to $Q_{sA}^2(b=0)/Q_{sp}^2(b=0) \sim 1$.

\end{abstract}
  
\section{Introduction}
\label{sec:guzey_intro}

Nuclear shadowing (NS) is a general phenomenon of high-energy scattering with nuclei, which consists in the observation that nuclear cross sections are 
smaller than the sum of nucleon cross sections. In the case of hard processes with nuclei, NS manifests itself as a suppression of nuclear structure functions and parton distributions (PDFs) at small momentum fractions $x < 0.1$. Thus, NS is fundamental for the description of the nuclear structure in perturbative quantum chromodynamics (QCD) and 
affects such aspects of it as initial conditions (cold nuclear matter effects) in proton-nucleus and nucleus-nucleus scattering at the Relativistic Heavy Ion Collider (RHIC) and the Large Hadron Collider (LHC) as well as an onset of the non-linear parton dynamics of the gluon-rich nuclear matter in the color glass condensate (CGC) 
framework.

The cleanest way to probe NS is provided by nuclear deep inelastic scattering (DIS), whose results are traditionally presented in the form of the 
ratio of the nucleus and nucleon structure functions $F_{2A}(x,Q^2)/[A F_{2N}(x,Q^2)]$. However, the same fixed-target nuclear DIS data can be successfully 
explained by different, competing mechanisms of NS. These include leading twist NS encoded in nuclear PDFs obtained from global QCD analyses of the available data~\cite{Klasen:2023uqj}, nucleus-enhanced power (higher-twist) corrections~\cite{Qiu:2003vd}, and a mixture of leading twist and higher-twist effects in the dipole
model with gluon saturation~\cite{Kowalski:2007rw}. This raises the question of the dynamical mechanism of NS and its relation to parton saturation.

This outstanding question of small-$x$ QCD (and many more) will be addressed at the planned Electron-Ion Collider (EIC) in USA~\cite{Accardi:2012qut}, which has the potential to discriminate among various approaches to NS due to a wide $x-Q^2$ kinematic coverage, an access to the nuclear longitudinal structure function $F_{L}^A(x,Q^2)$, and for the
 first time measurement of hard diffraction in nuclear DIS. In the latter case, it has been suggested that the ratio of the diffractive-to-total DIS cross sections
 for a heavy nucleus and the proton $R_{\rm diff/tot}$ is a sensitive observable. Indeed, while $R_{\rm diff/tot} > 1$ because of a nuclear enhancement 
 of the saturation scale $Q^2_{s,A}$ in the dipole framework~\cite{Kowalski:2007rw,Lappi:2023frf}, strong leading twist NS predicts that
 $R_{\rm diff/tot} < 1$~\cite{Frankfurt:2011cs}.

\section{Leading twist approach to NS in diffractive DIS}
\label{sec:guzey_lta}

The leading twist approach (LTA) to NS of~\cite{Frankfurt:2011cs} is a method to calculate various nuclear PDFs (usual, generalized, diffractive), 
which can be used as input for their $Q^2$ Dokshitzer-Gribov-Lipatov-Altarelli-Parisi (DGLAP) evolution. It is based on the Gribov-Glauber model of NS for soft
hadron-nucleus scattering and the QCD collinear factorization theorems for inclusive and diffractive DIS.

Constructing the $\gamma^{\ast}+A \to X+A$ amplitude as a series of diffractive scattering off $i=1,2,\dots, A$ target nucleons, 
one obtains for the ratio of the nuclear and proton diffractive PDFs in the small-$x$ limit~\cite{Frankfurt:2003gx,Guzey:2024xpa},
\begin{equation}
\frac{f_{i/A}^{D(3)}(x,x_{\Pomeron},Q^2)}{f_{i/p}^{D(3)}(x,x_{\Pomeron},Q^2)}= \frac{1}{\sigma_{\rm el}^i(x)} \int d^2 \vec{b}\,\left|1-e^{-\frac{1-i \eta}{2} \sigma_{\rm soft}^i(x)T_A(\vec{b})}\right|^2 \,,
\label{eq:guzey_coh}
\end{equation}
where $T_A(\vec{b})=\int dz \rho_A(\vec{b},z)$ with $\rho_A(\vec{r})$ the nuclear density, and 
$\sigma_{\rm el}^i(x)=[\sigma_{\rm soft}^i(x)]^2/(16 \pi B_{\rm diff})$ with $B_{\rm diff} \approx 6$ GeV$^{-2}$ the slope of the 
$t$ dependence of the proton diffractive structure function. 
The magnitude of NS is controlled by its relation to diffraction on the nucleon through the effective cross section
$\sigma_{\rm soft}^i(x)$,
\begin{equation}
\sigma_{\rm soft}^i(x) =\frac{16 \pi}{f_{i/p}(x)} \int_{x}^{0.1} \frac{dx_{\Pomeron}}{x_{\Pomeron}} f_{i/p}^{D(4)}(x,x_{\Pomeron},t=0) \,.
\label{eq:sigma_soft}
\end{equation}
Alternatively, $\sigma_{\rm soft}^i(x)$ can be calculated using a model for the hadronic structure of virtual photons.
This variation of $\sigma_{\rm soft}^i(x)$ determines the range of predictions between the ``high shadowing'' and 
``low shadowing'' scenarios, which is shown by the shaded bands in the figures. 
Equation~(\ref{eq:guzey_coh}) has a transparent physical interpretation: nuclear diffractive PDFs are shadowed in proportion to the nuclear elastic cross section.

Using completeness of nuclear final states, one can derive a similar expression for nuclear diffractive PDFs probed in 
quasi-elastic (summed) nuclear DIS~\cite{Guzey:2024xpa},
\begin{eqnarray}
\frac{\tilde{f}_{i/A}^{D(3)}(x,x_{\Pomeron},Q^2)}{f_{i/p}^{D(3)}(x,x_{\Pomeron},Q^2)} &=&  \frac{1}{\sigma_{\rm el}^i(x)} \int d^2 \vec{b}\, \Big(\left|1-e^{-\frac{1-i \eta}{2} \sigma_{\rm soft}^i(x)T_A(\vec{b})}\right|^2 \nonumber\\
&+&e^{- \sigma_{\rm in}^i(x)T_A(\vec{b})} -e^{-\sigma_{\rm soft}^i(x)T_A(\vec{b})}\Big)\,,
\label{eq:guzey_coh2}
\end{eqnarray}
where $\sigma_{\rm in}^i(x)=\sigma_{\rm soft}^i(x)-\sigma_{\rm el}^i(x)$.
In Eq.~(\ref{eq:guzey_coh2}), NS is given by a sum of the elastic and inelastic nuclear cross sections.

\begin{figure}[t]
\centerline{%
\includegraphics[width=10.cm]{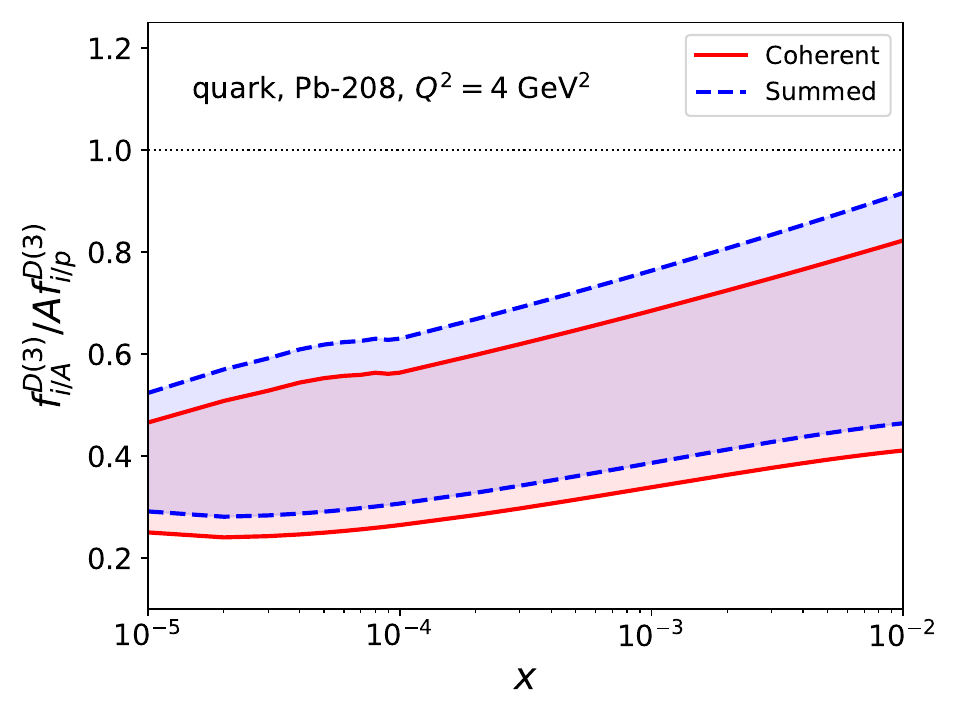}
}
\caption{The LTA predictions for the ratios of nucleus and proton diffractive PDFs, 
$f_{i/A}^{D(3)}/[Af^{D(3)}_{i/p}]$ and $\tilde{f}_{i/A}^{D(3)}/[Af^{D(3)}_{i/p}]$, as a function of $x$ at $Q^2=4$ GeV$^2$ for $^{208}$Pb.}
\label{fig:guzey_Shad_diffraction_quark}
\end{figure}

Figure~\ref{fig:guzey_Shad_diffraction_quark} shows the LTA predictions for the ratios of Eqs.~(\ref{eq:guzey_coh}) and (\ref{eq:guzey_coh2}) 
scaled down by factor $A$ for sea quarks as a function of $x$ at $Q^2=4$ GeV$^2$ for $^{208}$Pb. For the ratio of the gluon diffractive PDFs, the results are very similar.
One can see from this figure that LTA predicts that $f_{i/A}^{D(3)}/(Af^{D(3)}_{i/p}) \approx 
\tilde{f}_{i/A}^{D(3)}/(Af^{D(3)}_{i/p}) \approx F^{D(3)}_{2A}/(AF^{D(3)}_{2p})  \approx 0.5$ in a wide range of $x$
and independently of $x_{\Pomeron}$ provided that it is small. This suppression due to leading twist NS is very strong and should be compared to the impulse approximation prediction $f_{i/A}^{D(3)}/(Af^{D(3)}_{i/p})_{\rm IA}=4.3$.

Combing LTA predictions for diffractive and usual nuclear PDFs, one finds for the ratio $R_{\rm diff/tot}$ discussed in the Introduction, 
\begin{eqnarray}
 && R_{\rm diff/tot}= \frac{f_{i/A}^{D(3)}(x,x_{\Pomeron},Q^2)/f_{i/A}(x,Q^2)}{f_{i/p}^{D(3)}(x,x_{\Pomeron},Q^2)/f_{i/p}(x,Q^2)} 
= \frac{\sigma_{\rm soft}^i(x)}{\sigma^i_{\rm el}(x)} \nonumber\\
&\times& \frac{\int d^2 \vec{b} \left|1-e^{-\frac{1-i\eta}{2} \sigma_{\rm soft}^i(x) T_A(\vec{b})} \right|^2}
{2(1-\lambda^i(x)) \Re e \int d^2 \vec{b} \left(1-e^{- \frac{1-i\eta}{2} \sigma_{\rm soft}^i(x)T_A(\vec{b})}\right)+\lambda^i(x) A \sigma_{\rm soft}^i(x)} \,,
\label{eq:P_Ap}
\end{eqnarray}
where $\lambda^i(x)$ is the fraction of point-like (non-shadowed) configurations of the virtual photon.
Note that $\lambda^i(x)=0$, when $\sigma_{\rm soft}^i(x)$ is given by Eq.~(\ref{eq:sigma_soft}). A similar expression can be presented in the case of quasi-elastic diffraction, see Eq.~(\ref{eq:guzey_coh2}) and Ref.~\cite{Guzey:2024xpa}.

\begin{figure}[t]
\centerline{%
\includegraphics[width=10.cm]{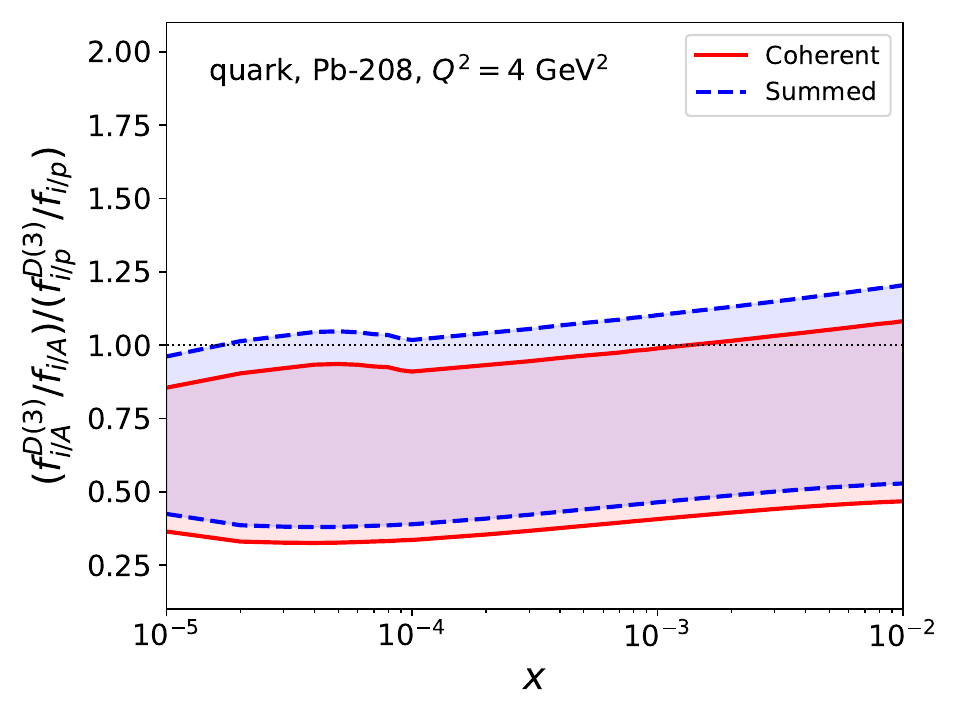}}
\caption{The LTA predictions for $R_{\rm diff/tot}$ of Eq.~(\ref{eq:P_Ap})
for sea quarks as a function of $x$ at $Q^2=4$ GeV$^2$ for $^{208}$Pb. The upper and lower curves correspond to the ``high shadowing'' and ``low shadowing'' scenarios.}
\label{fig:guzey_Shad_diffraction_PAp_quark}
\end{figure}

Figure~\ref{fig:guzey_Shad_diffraction_PAp_quark} shows the LTA predictions for the ratio $R_{\rm diff/tot}$ of Eq.~(\ref{eq:P_Ap})
for sea quarks as a function of $x$ at $Q^2=4$ GeV$^2$ for $^{208}$Pb.
The shaded bands quantity the significant LTA theoretical uncertainties for
this ratio, where the upper and lower curves correspond to the ``high shadowing'' and ``low shadowing'' scenarios.
One can see from the figure that $R_{\rm diff/tot} \approx 0.5-1$ for sea quarks, which reaffirms the earlier LTA result~\cite{Frankfurt:2011cs} and its difference from the nuclear enhancement of $R_{\rm diff/tot}$ expected in the 
gluon saturation framework~\cite{Kowalski:2007rw,Lappi:2023frf}. 
For the ratio of the corresponding gluon PDFs, $R_{\rm diff/tot} \approx 0.5-1.3$.
An inspection shows that this results from an interplay between large leading twist NS for diffractive and usual nuclear PDFs.

To better understand these results and compare them with the dipole model, one can examine $R_{\rm diff/tot}$ as function of 
$\sigma_{\rm soft}^i(x)$ and $\lambda^i(x)$. The values below refer to the summed case. 
(i) For small $\sigma_{\rm soft}^i(x)$, NS is vanishingly small and $R_{\rm diff/tot} \approx 5$. 
(ii) As $\sigma_{\rm soft}^i(x)$ increases and reaches $\sigma_{\rm soft}^i(x) \sim \sigma_{\rho N} \approx 20-30$ mb, 
the corresponding NS is still weak with $R_{\rm diff/tot} \approx 1.5-2$. It corresponds to the situation encountered in the dipole model.
(iii) When $\sigma_{\rm soft}^i(x) \geq 40$ mb, one reaches the regime of the full-fledged leading twist NS, which predicts that 
$R_{\rm diff/tot} \approx 0.5-1.3$. This large spread is caused by the strong dependence on the value of $\lambda^i(x)$.

Note that $\sigma_{\rm soft}^i(x)$ cannot grow forever because it is bounded by the black disk limit (BDL): 
$\sigma_{\rm soft}^i(x) \leq 8 \pi B_{\rm diff} \approx 60$ mb. In this case, all hadronic configuration of the virtual photon interact 
with a nuclear target with the same maximal cross section and, hence, $\lambda^i(x)=0$, and one finds that $R_{\rm diff/tot} \to 1$ in the summed case and
$R_{\rm diff/tot} \to 0.86$ in the case of purely coherent scattering.

\section{Leading twist NS and the saturation scale}
\label{sec:guzey_Qs}

The saturation scale $Q_s^2$ can be heuristically defined through the impact parameter dependent gluon density $g(x,b,Q^2)$, where the impact parameter 
$\vec{b}$ denotes the distance from the nucleus center.  Then
the ratio of the saturation scales for a heavy nucleus and the proton is~\cite{Guzey:2024xpa,Frankfurt:2022jns}
\begin{eqnarray}
\frac{Q_{sA}^2(b)}{Q_{sp}^2(b)}&=&\frac{g_A(x,b,Q^2)}{g_p(x,b,Q^2)}=\pi R_p^2 \,\frac{g_A(x,b,Q^2)}{g_p(x,Q^2)}=\pi R_p^2  \nonumber\\&\times& \left[\lambda^i(x)T_A(\vec{b}) +\frac{2(1-\lambda^i(x))}{\sigma_{\rm soft}^i(x)} \Re e \left(1-e^{-\frac{1-i\eta}{2} \sigma_{\rm soft}^i(x) T_A(\vec{b})}\right)\right] \,,
\label{eq:guzey_Q_sat}
\end{eqnarray}
where we used the Gaussian $|\vec{b}|$-profile for the proton distribution, $g_p(x,b,Q^2)=e^{-b^2/R_p^2}/(\pi R_p^2) g_p(x,Q^2)$ with
 $R_p^2=2 B_{J/\psi}=9+0.4 \ln (x_0/x)$ GeV$^{-2}$. The use of Eq.~(\ref{eq:guzey_Q_sat}) at $\vec{b}=0$ gives
 \begin{equation}
\frac{Q_{sA}^2(b=0)}{Q_{sp}^2(b=0)} \approx \frac{2 \pi R_p^2}{\sigma_{\rm soft}^i(x)} \sim 1 \,.
\label{eq:guzey_Q_sat2}
\end{equation}
It should be contrasted with the impulse approximation prediction $Q_{sA}^2(b=0)/Q_{sp}^2(b=0)_{\rm IA}= \pi R_p^2 T_A(b=0) \sim A^{1/3}$. 
The estimate of Eq.~(\ref{eq:guzey_Q_sat2}) is a result of the strong leading twist NS and the dilute realistic nuclear density.

\section{Conclusions}
\label{sec:guzey_conclusions}

The dynamical mechanism of NS and its relation to parton saturation is one of outstanding questions of QCD at high energies.
It has been suggested 
that the ratio of the diffractive-to-total DIS cross sections $R_{\rm diff/tot}$ for a heavy nucleus and proton
at the EIC has the potential to discriminate between the leading twist and saturation-based approaches to NS. 
In this study we confirmed that $R_{\rm diff/tot} \approx 0.5-1$ for sea quarks and $R_{\rm diff/tot} \approx 0.5-1.3$ for gluons in LTA in contrast with $R_{\rm diff/tot} \approx 1.5-2$ in the gluon saturation framework. 
We showed that $R_{\rm diff/tot}$ is controlled by an effective (dipole) cross section, which is large in LTA due to its
connection to diffraction on proton and small in the dipole model. In particular, $R_{\rm diff/tot} \to 1$ in the black disk limit.
Finally, we argued that strong leading twist NS as well as the dilute nuclear density 
deplete nuclear enhancement of the saturation scale leading to $Q_{sA}^2(b=0)/Q_{sp}^2(b=0) \sim 1$.

{\small This research was funded by the Research Council of Finland, the Centre of Excellence in Quark Matter (projects 364191 and 364194), and the European Research Council project ERC-2018-ADG-835105 YoctoLHC. }

\end{document}